\documentstyle[11pt,aaspp4]{article}
\begin{document}
\newcommand{\lya}{Lyman~$\alpha$}
\newcommand{\lyb}{Lyman~$\beta$}
\newcommand{\za}{$z_{\rm abs}$}
\newcommand{\ze}{$z_{\rm em}$}
\newcommand{\cmtwo}{cm$^{-2}$}
\newcommand{\nhi}{$N$(H$^0$)}
\newcommand{\nzn}{$N$(Zn$^+$)}
\newcommand{\ncr}{$N$(Cr$^+$)}
\newcommand{\degpoint}{\mbox{$^\circ\mskip-7.0mu.\,$}}
\newcommand{\halpha}{\mbox{H$\alpha$}}
\newcommand{\hbeta}{\mbox{H$\beta$}}
\newcommand{\hgamma}{\mbox{H$\gamma$}}
\newcommand{\kms}{\,km~s$^{-1}$}      
\newcommand{\minpoint}{\mbox{$'\mskip-4.7mu.\mskip0.8mu$}}
\newcommand{\mv}{\mbox{$m_{_V}$}}
\newcommand{\Mv}{\mbox{$M_{_V}$}}
\newcommand{\peryr}{\mbox{$\>\rm yr^{-1}$}}
\newcommand{\secpoint}{\mbox{$''\mskip-7.6mu.\,$}}
\newcommand{\sqdeg}{\mbox{${\rm deg}^2$}}
\newcommand{\squig}{\sim\!\!}
\newcommand{\subsun}{\mbox{$_{\twelvesy\odot}$}}
\newcommand{\et}{{\it et al.}~}

\def\ltsima{$\; \buildrel < \over \sim \;$}
\def\simlt{\lower.5ex\hbox{\ltsima}}
\def\gtsima{$\; \buildrel > \over \sim \;$}
\def\simgt{\lower.5ex\hbox{\gtsima}}
\def\arcs{$''~$}
\def\arcm{$'~$}
\def\erf{\mathop{\rm erf}}
\def\erfc{\mathop{\rm erfc}}
\title{A COUNTS-IN-CELLS ANALYSIS OF LYMAN-BREAK GALAXIES AT REDSHIFT $Z\sim 3$\altaffilmark{1}}
\author{\sc Kurt L. Adelberger and Charles C. Steidel\altaffilmark{2,3}}
\affil{Palomar Observatory, Caltech 105--24, Pasadena, CA 91125}
\author{\sc Mauro Giavalisco\altaffilmark{4}}
\affil{The Carnegie Observatories, 813 Santa Barbara Street, Pasadena, CA 91101}
\author{\sc Mark Dickinson\altaffilmark{5,6}}
\affil{Department of Physics and Astronomy, The Johns Hopkins University, Baltimore, MD 21218}
\author{\sc Max Pettini}
\affil{Royal Greenwich Observatory, Madingley Road, Cambridge CB3 0EZ, UK}
\author{\sc Melinda Kellogg}
\affil{Palomar Observatory, Caltech 105--24, Pasadena, CA 91125}

\altaffiltext{1}{Based in part on observations obtained at the W.M. Keck
Observatory, which is operated jointly by the California Institute of
Technology and the University of California.} 
\altaffiltext{2}{Alfred P. Sloan Foundation Fellow}
\altaffiltext{3}{NSF Young Investigator}
\altaffiltext{4}{Hubble Fellow}
\altaffiltext{5}{Alan C. Davis Fellow}
\altaffiltext{6}{also Space Telescope Science Institute, 3700 San Martin Drive, Baltimore, MD 21218}
\begin{abstract}
We have measured the counts-in-cells fluctuations of 268 Lyman-break
galaxies with spectroscopic redshifts
in six 9\arcm$\times$ 9\arcm fields at $z\sim 3$.  The
variance of galaxy counts in cubes of comoving
side length 7.7, 11.9, 11.4 $h_{100}^{-1}$ Mpc
is $\sigma_{\rm gal}^2 \sim 1.3\pm0.4$ for $\Omega_M=1$, $0.2$~open,
$0.3$~flat, implying
a bias on these scales of
$\sigma_{\rm gal} / \sigma_{\rm mass} = 6.0\pm1.1$, $1.9\pm0.4$, $4.0\pm0.7$.
The bias and abundance
of Lyman-break galaxies are surprisingly consistent with a simple model
of structure formation which assumes only that galaxies form
within dark matter halos, that Lyman-break galaxies' rest-UV luminosities
are tightly correlated with their dark masses, and that matter
fluctuations are Gaussian and have a linear power-spectrum shape
at $z\sim 3$ similar to that
determined locally ($\Gamma\sim 0.2$).  This conclusion is largely
independent of cosmology or spectral normalization $\sigma_8$.
A measurement of the masses of Lyman-break galaxies would in
principle distinguish between different cosmological scenarios.
\end{abstract}

\section{INTRODUCTION}
Much of observational cosmology depends upon the assumption that
the spatial distribution of galaxies is related in a simple way
to the underlying distribution of matter.  At first
it was hoped that the galaxy distribution might simply be
a Poisson realization of the matter distribution; but as
this model became difficult to reconcile with large scale
peculiar velocities, the amplitude of microwave background
fluctuations, the different clustering strengths of different
galaxy types, and theoretical prejudice for $\Omega_M=1$, 
cosmologists began to assume an unspecified constant
of proportionality $b$ between galaxy and mass fluctuations:
$\delta_{\rm gal} = b \delta_{\rm mass}$.  Though many physical
processes could in principle give rise to a relationship of
this form (e.g. Dekel \& Rees 1987), most were
poorly understood, and, if invoked, would 
make it difficult to use galaxy observations to constrain
the cosmological mass distribution.  An important exception
was gravitational instability.  This is relatively well
understood, and if it were dominant in determining where
galaxies formed---if galaxies formed within virialized
``halos'' of dark matter, and if the poorly understood
physics of star formation, supernova feedback, and so on were
important only in determining the properties of galaxies within
dark matter halos---then the large scale distribution of galaxies
would still be related in a simple way to the underlying
distribution of matter; the value of the ``bias parameter'' $b$ in
$\delta_{\rm gal} = b \delta_{\rm mass}$ would be straightforward
to calculate (White \& Rees 1978, Kaiser 1984, Bardeen \et 1986,
Mo \& White 1996).  Because it maintains a simple relationship
between galaxies and mass, agrees with our limited knowledge
of the relevant physics, and seems consistent with
numerical simulations, this ``dark halo'' model has become
increasingly popular, and is now the basis of the modern
understanding of galaxy formation.  It is 
assumed in most analytic treatments, in semi-analytic models,
and in numerical simulations which include only gravity; and yet
it remains a conjecture that has never been thoroughly tested.
One prediction of
the dark halo model is that galaxies of a given mass should form
first in regions where the density is highest, and since such regions
are expected to be strongly clustered (e.g. Kaiser 1984), a natural
test is to measure the clustering of galaxies in the young universe.

The Lyman-break technique (e.g., Steidel, Pettini, \& Hamilton 1995) 
provides a way to find large numbers of
star-forming galaxies at $z\sim 3$.  Star-forming galaxies have
pronounced breaks in their spectra at 912 \AA\ (rest)
from a combination of absorption by neutral hydrogen
in their interstellar media and the intrinsic spectra
of massive stars.
At $z\simgt 3$ this ``Lyman
break,'' strengthened from additional absorption by hydrogen in the
unevolved intergalactic medium, is redshifted sufficiently to
be observed with ground-based broad-band photometry.  By taking
images through filters that straddle the redshifted Lyman break,
and looking for objects that are much fainter
in images at wavelengths shortward of the break
than longward of the break, one can
efficiently separate high-redshift galaxies from
the many foreground objects.
In our implementation of the technique, we have used deep
photometry in the custom $U_n$, $G$, ${\cal R}$ filter system
of Steidel \& Hamilton (1993) to assemble a sample of
over~1300 probable $z\sim 3$ galaxies, of which more
than~400 have been spectroscopically confirmed with the
Low-Resolution Imaging Spectrograph (Oke \et 1995) on the W.~M.~Keck
telescopes.
            
After initial spectroscopy in one 9\arcm$\times$ 18\arcm field, we argued,
on the basis of a single large concentration of galaxies, that
these $z\sim 3$ Lyman-break galaxies were much more strongly clustered
than the mass, with an inferred bias parameter
$b\equiv \sigma_{\rm gal}/\sigma_{\rm mass}$
of $b\simgt 6$, $2$, $4$ for $\Omega_M=1$, $0.2$~open, and $0.3$~flat
(Steidel \et 1998a).
Qualitatively this strong biasing was consistent with the idea
that galaxies form first in the (strongly clustered) densest regions of
the universe, but there appeared to be quantitative problems.
In the dark matter halo model there is an inverse relationship between
the abundance and bias of a population of halos, with the
rarest, most massive halos being the most strongly clustered
(i.e., most ``biased'').
As emphasized by Jing \& Suto (1998), for halos 
to be as strongly clustered
as Lyman-break galaxies, they would have to be very rare indeed.
Yet Lyman-break galaxies are not that rare;
for $\Omega_M=1$ their comoving
number density to ${\cal R}=25.5$
is $n\simgt 8\times 10^{-3}$ per $h_{100}^{-3}$ Mpc$^3$, comparable to the
number density of $L_*$ galaxies today.  As we will see below,
in standard ($\Omega_M=1$, $\sigma_8=0.6$, $\Gamma=0.50$) CDM,
halos at $z\sim 3$ with the same abundance as observed Lyman-break
galaxies have a bias of $b\sim 4$,
substantially lower than the implied galaxy bias.
For $\Omega_M<1$ the disagreement is less severe, because
both the estimated bias and the comoving abundance of observed Lyman-break
galaxies are lower.  It appeared then from preliminary analyses that
our data were consistent with the dark halo model only
for $\Omega_M<1$; but it was unclear how seriously to take
conclusions based on a single feature in a single field.
Moreover other authors soon analyzed the overdensity differently and
argued that it was consistent with models in which galaxies are
significantly less clustered than we claimed, with $b$ low enough
to remove the inconsistencies with the abundances for $\Omega_M=1$
(e.g. Bagla 1997, Governato \et 1998, Wechsler \et 1998).

In this paper we present a counts-in-cell analysis of the clustering
of 268 Lyman-break galaxies (all with spectroscopic redshifts)
in six 9\arcm$\times$ 9\arcm fields.
This sample contains
four times as many galaxies over an area three times as large as
our original analysis.  Since in addition it takes into account
all galaxy fluctuations in the data, and not just a single over-density,
one might hope it would provide a more definitive measurement of
the strength of clustering.

\section{DATA}
Many relevant details of our survey for Lyman-break galaxies
are presented elsewhere (Steidel \et 1996, Giavalisco \et 1998a,
Steidel \et 1998a, Steidel \et 1998b),
and in this section we give only a brief review.
We initially identify $z\sim 3$ galaxy candidates in
deep $U_n$, $G$, ${\cal R}$ images taken (primarily) at the Palomar 5m Hale
telescope with the COSMIC prime focus camera.
In images of our typical depths (1$\sigma$ surface brightness
limits of 29.1, 29.2, 28.6 AB magnitudes per arcsec$^2$
in $U_n$,$G$, and ${\cal R}$)
approximately 1.25 objects per arcmin$^2$ meet our current selection criteria of
$${\cal R}\le 25.5,\quad G-{\cal R}\le 1.2,\quad U_n-G\ge G-{\cal R}+1,\quad U_n-G\ge 1.6.$$
A subset of these photometric candidates is subsequently observed
spectroscopically at the W.~M.~Keck telescope through multislit masks
which accommodate $\sim 20$ objects each.
To date we have obtained spectra of 540 objects satisfying the
above photometric criteria; 376 of these have been identified as
galaxies (of which a very small fraction show evidence of AGN activity),
with a redshift distribution shown in Figure~1; 18 are
stars; and the remainder have not been identified because of
inadequate signal to noise ratio.
In this paper we restrict our analysis to the 268 Lyman-break
galaxies in our six most densely sampled $\sim$~9\arcm$\times~$9\arcm\
fields, including more complete data in the ``SSA22'' field
analyzed in Steidel \et 1998a.  The redshift histograms 
these six fields are shown in Figure~2; each field is treated independently in
the analysis that follows, although in two cases (SSA22 and DSF2237) pairs
of 9\arcm fields are adjacent on the plane of the sky.

\section{STATISTICAL ANALYSIS}
The strength of clustering can be estimated by placing galaxies into
spatial bins (``cells'') and looking at the fluctuations in galaxy
counts from cell to cell.  A convenient measure of the clustering
strength is 
$$\sigma_{\rm gal}^2 \equiv {1\over V_{\rm cell}^2} \int\int_{V_{\rm cell}}dV_1dV_2\xi_g(r_{12})$$
where $\xi_g(r)$ is the galaxies' two-point correlation function.  If there
were large numbers of galaxies in each cell, so that shot noise were
negligible, $\sigma_{\rm gal}^2$ would just be equal to the
relative variance of galaxy counts in cells of volume $V_{\rm cell}$:
$\sigma_{\rm gal}^2 = <(N-\mu)^2>/\mu^2$, where $N$ is the observed 
and $\mu$ the mean number of galaxies in a cell.  In practice
shot noise makes a significant contribution to the variance of
cell counts, and this contribution must be removed to estimate
$\sigma_{\rm gal}^2$:
$$\sigma_{\rm gal}^2 = (<(N-\mu)^2>-\mu)/\mu^2$$
(Peebles 1980, \S 36).
For any cell the expected number of galaxies can be estimated accurately
as $\mu\simeq N_{\rm tot} \phi(z) \Delta z$, where $\phi(z)$ is our
selection function, determined by fitting a spline to
the coarsely binned redshifts of all $\sim 400$ Lyman-break galaxies
which satisfy our current color criteria and have redshifts,
and $N_{\rm tot}$ is the number of galaxies in the
field with redshifts.  ($N_{\rm tot}$ varies from field to field because
of differing spectroscopic completeness.)
In general the uncertainty in cell count $N$ will dominate the uncertainty
in $\mu$.  If we neglect the relatively small uncertainty in $\mu$,
we can estimate
$\sigma_{\rm gal}^2$ from the number of counts $N$ in a single cell as
$${\cal S} = ((N-\mu)^2-\mu)/\mu^2.$$
If $\mu$ were perfectly known, ${\cal S}$ would have expectation value
$<{\cal S}> = \sigma_{\rm gal}^2$ and variance
$$ <{\cal S}^2> - <{\cal S}>^2 = 2\sigma_{\rm gal}^4 + 4\sigma_{\rm gal}^2/\mu + (2+7\sigma_{\rm gal}^2)/\mu^2 + 1/\mu^3\eqno(1)$$
where we have used results in Peebles (1980, \S 36) and neglected
the integrals over the three- and four-point correlation functions.  In fact
${\cal S}$ will be a slightly biased estimator of $\sigma_{\rm gal}^2$, since
our estimate of $\mu$ depends weakly on $N$ (through its contribution to $N_{\rm tot}$),
but this bias should be small compared to the variance---which is itself only
approximately equal to the RHS of equation 1.  
With ${\cal S}$ we can estimate $\sigma_{\rm gal}^2$ from the
observed number of counts in a single cell; by combining the
estimates ${\cal S}$ from every cell in our data with inverse-variance
weighting, we arrive at our best estimate of $\sigma_{\rm gal}^2$.
(The variance depends on the unknown $\sigma_{\rm gal}^2$, of course,
but the answer converges with a small number of iterations.)

Placing our counts into a dense grid of roughly cubical cells whose
transverse size is equal to the field of view ($\sim$ 9\arcm), we
estimate $\sigma_{\rm gal}^2 = 1.3\pm 0.4$ in cells
of approximate length 7.7, 11.9, 11.4 $h_{100}^{-1}$ Mpc for 
$\Omega_M=1$, $0.2\,{\rm open}$, $0.3\,{\rm flat}$.
The uncertainty is the standard deviation of the mean of ${\cal S}$
estimated in the fields individually.

This approach with the estimator ${\cal S}$ has the advantage of
being relatively model independent, but
statisticians have long argued that an optimal data analysis must
use the likelihood function (e.g. Birnbaum 1962).
If we had a plausible model for the probability density function (PDF)
of galaxy fluctuations $P(\delta_{\rm gal}|\sigma_{\rm gal}^2)$,
we might hope to produce a better estimate of $\sigma_{\rm gal}^2$
by finding the value that maximizes the likelihood of the data.
An exact expression for the galaxy PDF has not been found, but
it should be sufficient to use a reasonable approximation.
The main requirement for this approximate PDF is that it be skewed,
since a galaxy fluctuation
$\delta_{\rm gal}\equiv(\rho_{\rm gal} - \bar\rho_{\rm gal})/\bar\rho_{\rm gal}$
can be arbitrarily large but cannot be less than -1.  A particularly
simple distribution with the necessary skew, the lognormal,
provides a good fit to the PDF of mass fluctuations
and of linearly biased galaxies in N-body simulations
(e.g. Coles \& Jones 1991, Coles \& Frenk 1991, Kofman \et 1994).
The lognormal probability of observing a galaxy fluctuation $\delta_{\rm gal}$
given $\sigma_{\rm gal}^2$ is
$$P_{LN}(\delta_{\rm gal}|\sigma_{\rm gal}^2) =
{1\over 2\pi x}\exp\biggl[-{1\over 2}\biggl({\log(1+\delta_{\rm gal})\over x} +
{x\over 2}\biggr)^2\biggl]$$
where $x\equiv\sqrt{\log(1+\sigma_{\rm gal}^2)}$, and so in this
model, assuming Poisson sampling, the likelihood of observing $N$ galaxies
in a cell when $\mu$ are expected is
$$P(N|\mu\sigma_{\rm gal}^2) = \int_{-1}^{\infty}d\delta_{\rm gal} P_{LN}(\delta_{\rm gal}|\sigma_{\rm gal}^2) \exp[-(1+\delta_{\rm gal})N] (1+\delta_{\rm gal})^N\mu^N / N!.$$
The analytical solution to this integral is unknown, but it presents no
numerical challenge.
If the cells are large enough to be nearly uncorrelated, we can find
the maximum likelihood value of $\sigma_{\rm gal}^2$ by maximizing
the product of the likelihoods from individual cells. 
Figure~3 shows the product of the likelihoods for $\sigma_{\rm gal}^2$ from
all cells in all six fields for $\Omega_M=1$.  The plots for
$\Omega_M=0.2\,{\rm open}$ and $\Omega_M=0.3\,{\rm flat}$ are similar,
with small differences arising because our desire for cubical cells
forces us to use different redshift binning for different cosmologies.
For each cosmology the overall maximum likelihood value
is close to $\sigma_{\rm gal}^2\sim 1.3$; the 68.3\% credible intervals
are 0.8 to 1.6, 0.7 to 1.4, and 1.1 to 2.1 for $\Omega_M=1$, 
$\Omega_M=0.2$~open, and $\Omega_M=0.3$~flat,
in reasonable agreement with our estimate from ${\cal S}$.\footnote{This
approach to estimating $\sigma_{\rm gal}^2$ is very similar to
that of Peacock (1997).}  We will take the maximum likelihood estimates
as our best estimates of $\sigma_{\rm gal}^2$ hereafter.

A more common measure of the clustering strength is the characteristic length
$r_0$ in a correlation function of assumed form $\xi_g(r)=(r/r_0)^{-\gamma}$.
For spherical cells $r_0$ and $\sigma_{\rm gal}^2$ are related through
$\sigma_{\rm gal}^2 = 72(r_0/R_{\rm cell})^\gamma / [(3-\gamma)(4-\gamma)(6-\gamma)2^\gamma]$ (Peebles 1980 \S 59),
and so approximating our cubical cells as spheres with equal volume,
and assuming $\gamma=1.8$, we arrive at a rough estimate of
$r_0\simeq 4\pm 1$, $5\pm 1$, $6\pm 1$ comoving $h_{100}^{-1}$ Mpc
for $\Omega_M=1$, $0.2$~open, $0.3$~flat.  These values are large,
comparable to the correlation lengths of massive galaxies today.

The correlation lengths for $\Omega_M=1$ and $\Omega_M=0.2$~open
are larger, by about $2\sigma$, than those recently derived
by Giavalisco \et (1998a, ``G98a'' hereafter) from the angular clustering of
Lyman-break galaxies.  This discrepancy could be resolved
in several ways.  The correlation lengths would
agree at the $1\sigma$ level if a large fraction of
the objects whose spectra we cannot identify (about 25\% of
the spectroscopic sample) were lower redshift interlopers
diluting the angular clustering signal.  The discrepancy would also
be reduced if $\gamma$ were larger than 1.8, although $\gamma$
would have to be equal to $\sim 2.6$, contradicting the results
of G98a, to make the correlation lengths agree at the
$1\sigma$ level.  Because the spectroscopic subsample is
somewhat brighter than the sample as a whole, one would
expect (from arguments we develop below) the galaxies
analyzed here to be somewhat more strongly clustered than
those analyzed in G98a, but this would change $r_0$
by only $\sim$~10-20\% (these numbers follow from the formalism
presented below, and will be explained more fully in Giavalisco \et (1998b)).
Inferring $r_0$ from observed angular
clustering depends upon the assumed cosmological
geometry, because (for example) projection effects
must be corrected, and an intriguing possibility is that
the correlation lengths disagree because G98a assumed an incorrect
geometry when deriving $r_0$ from the angular clustering.
According to G98a, the quantity $A^{1/\gamma}$ (for
a correlation function of the assumed form
$\omega(\theta)=A\theta^{1-\gamma}$)
is well constrained by their observations.  If we take $A^{1/\gamma}$
and $\sigma_{\rm gal}^2$ as two cosmology-independent parameters
fixed by observation (which is not quite true; see above),
then the correlation length derived from $\omega(\theta)$ scales
with cosmological parameters roughly as
$r_{0,\omega}\propto(g(\bar z)/f^{1-\gamma}(\bar z))^{1/\gamma}$,
where $g\equiv dl/dz$ is the change in proper distance with redshift
and $f$ is the angular diameter distance, while
the correlation length derived from $\sigma_{\rm gal}^2$ 
roughly obeys $r_{0,\sigma^2}\propto f(\bar z)$.
The ratio of these correlation lengths therefore depends
on the geometry as $r_{0,\omega}/r_{0,\sigma^2}\propto (g/f)^{1/\gamma}$,
and so if we assume $f=f_f$ and $g=g_f$ when the correct values
are $f=f_t$ and $g=g_t$, we will find correlation lengths from
counts-in-cell and $\omega(\theta)$ analyses which differ
by a factor $\eta\equiv r_{0,\omega}/r_{0,\sigma^2}=(f_tg_f/f_fg_t)^{1/\gamma}$.
For $\gamma=1.8$ in an $\Omega_M=0.2$~flat cosmology, we would
find $\eta=0.88$ if we mistakenly assumed $\Omega_M=0.2$~open,
and $\eta=0.84$ if we assumed $\Omega_M=1$.
This does not go far towards reconciling the discrepant
correlation lengths, but it does suggest an interesting variant of
Alcock \& Paczynski's (1979) classic cosmological test.
Finally, G98a found differences of 30\% in $r_0$ when measuring
the angular clustering with different estimators, and this implies
that the systematics in that sample may not be fully understood.
While these effects taken together could easily reconcile the
results presented here with those of G98a, the differences
are significant and will likely only be convincingly resolved
by better data.  Because the largest corrections we have proposed
apply to the estimates of $r_0$ from angular clustering, we will take
the counts-in-cell result as our best estimate of the clustering
strength in our subsequent discussion.

\section{THE BIAS AND ABUNDANCE OF LYMAN-BREAK GALAXIES}

A large bias for high-redshift galaxies is a prediction of
models that associate galaxies with virialized dark matter halos
(e.g. Cole \& Kaiser 1989), and on the face of it the strong
clustering of Lyman-break galaxies seems
a significant success for them.  But 
these models explain strong clustering by associating
high-redshift galaxies with rare events in the underlying Lagrangian
density field, and would be ruled out if Lyman-break galaxies were
too common to be so strongly clustered.
In this section we examine the consistency
of clustering strength and abundance
in more detail; but before we can do so we need to estimate
the Lyman-break galaxies' bias.
Our definition of bias is the ratio of
rms galaxy fluctuation to rms mass fluctuation in cells of our chosen size:
$b\equiv\sigma_{\rm gal}/\sigma_{\rm mass}$.  The mean square
mass fluctuation in a cell at $z\sim 3$ can be calculated
with a numerical integration:
$\sigma_{\rm mass}^2 = (2\pi)^{-3} \int d^3k |\delta_k|^2 |W_k|^2$
(e.g., Padmanabhan 1993), where $W_k$ is the Fourier transform
of the cell volume and $|\delta_k|^2$ is the power-spectrum of
density fluctuations.  By most accounts the shape of the power-spectrum
is close to that of a CDM-like model with
``shape parameter'' $\Gamma\sim 0.2$ (Vogeley \et 1992, Peacock \& Dodds 1994,
Maddox \et 1996; we use Bardeen \et 1986 equations G2 and G3
with $q=k/\Gamma h$ and an $n=1$ long-wavelength limit 
as an approximation to the spectral shape).  The normalization of the
power-spectrum can be determined at $z=0$ from the
abundance of X-ray clusters, and on large scales of interest
here can be reliably extrapolated back to $z=3$
with linear theory.

One complication prevents us from simply dividing our
measured $\sigma_{\rm gal}$ by the calculated $\sigma_{\rm mass}$
to estimate the bias: 
we have measured the relative variance of galaxy counts in cells
defined in redshift space, and
this variance is boosted relative to the
real-space galaxy variance of interest by coherent infall
towards overdensities, and reduced by redshift measurement
errors.\footnote{We are assuming that these errors
dominate the pair-wise velocity dispersion (``finger of god'' effect).  A large
pairwise velocity dispersion decreases the size of redshift-space fluctuations
for a fixed size of real-space fluctuations, and so by neglecting the
dispersion we will underestimate the bias.  But the effect is not large;
the pairwise velocity dispersion would have to be $\sim 800$ km/s
to change our estimated bias by $1\sigma$.}
Both effects must be corrected before we can estimate
the bias.  Fortunately neither effect is large for
highly biased galaxies in cells of this size, and the correction
is straightforward.
Following Peacock \& Dodds (1994), we
estimate $b$ by numerically inverting
$$\sigma_{\rm gal}^2 = {b^2\over (2\pi)^3}\int d^3k |\delta_k|^2 |W_k|^2 (1+fk_z^2/k^2b)^2 \exp(-k_z^2\sigma_v^2)$$
where $f\simeq\Omega_M^{0.6}(z)$ and
$\sigma_v\simeq 300{{\rm km}\over{\rm s}}(1+z)/H(z)$ is the adopted
uncertainty in a galaxy's position from redshift measurement errors
(see Steidel \et 1998a).  This expression is a modified
version of the usual integral relationship between the variance
and the power-spectrum; the factor of $1+fk_z^2/k^2b$ in the
integrand accounts for the increase in redshift-space power (relative
to real-space power)
due to coherent infall (e.g. Kaiser 1987),
and the Gaussian models our redshift uncertainties.  Corrections
for the non-linear growth of perturbations on scales much smaller
than our cell (described, for example, in the same Peacock \& Dodds
reference) have been neglected.
The results of this bias calculation are shown in Figure~4.
With $\Gamma=0.2$ we find $b=6.0\pm 1.1$, $1.9\pm 0.4$, and
$4.0\pm 0.7$ for $\Omega_M=1$, $0.2$~open, and $0.3$~flat.
This estimate of the bias is inversely
proportional to the somewhat-uncertain power-spectrum normalization $\sigma_8$;
for concreteness we have chosen
$\sigma_8=0.5$, $1.0$, $0.9$ for $\Omega_M=1$, $0.2$~open, and $0.3$~flat,
close to the cluster normalization of Eke, Cole, \& Frenk (1996), but
our most important conclusions below, about the bias/abundance relationship,
are insensitive to the normalization.
Varying the spectral shape over the plausible range $\Gamma=\,$~0.1--0.5
changes our estimate of the
bias by about $\pm$~10\% for $\Omega_M=1$ and by a negligible
amount for the other cosmologies (see Fig. 4), assuming the $\Gamma$
dependence of $\sigma_8$ is negligible (e.g. White, Efstathiou, \& Frenk 1993).

We can test the idea that Lyman-break galaxies
form within dark matter halos by comparing their inferred bias to
the predicted bias of dark matter halos with similar abundance.
Simple statistical arguments (e.g. Kaiser 1984, Mo \& White 1996) show
that the main factor controlling the clustering strength of a population
of halos with mass $M$ is their ``rareness'' $\nu\equiv\delta_c/\sigma(M)$,
where $\sigma(M)$ is the rms
relative mass fluctuation in the density field smoothed by
a spherical top-hat enclosing average mass $M$, and $\delta_c\simeq 1.7$
is the linear overdensity corresponding to spherical collapse.  To first
order,
$$b\simeq 1+(\nu^2-1)/\delta_c\eqno(2)$$
(Mo \& White 1996).
The abundance of these same halos is approximately given 
by the Press-Schechter (1974) formula 
$$n(\nu)d\nu = {\bar\rho \over M(\nu,\Gamma)} \sqrt{2\over\pi} \exp(-\nu^2/2)d\nu\eqno(3)$$
where we have written the halo mass as $M(\nu,\Gamma)$ to emphasize that
it depends upon both the halos' rareness and the shape of the matter
power-spectrum, as discussed below.
This relation is easy to understand: $\bar\rho/M$ is the maximum possible
number density of collapsed objects on mass scale $M$, given the
finite average density of the universe $\bar\rho$,
and $\sqrt{2/\pi}\exp(-\nu^2/2)$---which
follows from the assumed Gaussian distribution of the linear
density field---is the fraction of this maximum number that has just reached
the threshold for collapse.  From the Press-Schechter formula is it clear
that the clustering strength of a population of given abundance will depend
upon the shape of the power-spectrum:  if the fluctuation spectrum has 
more small-scale power, then the process of collapse will have advanced
to larger mass scales,
$\bar\rho/M$ will be smaller, and $\nu$ (and therefore $b$, by equation 2)
will also have to be smaller to match the observed abundance.

If we were free to specify the shape of the fluctuation spectrum,
then, we could (almost) always
argue that our observation of a galaxy population with abundance $n$ and
bias $b$ was consistent with the idea that galaxies form within
dark matter halos, by simply adjusting the level of small-scale power
until halos of abundance $n$ were predicted to have
bias $b$;\footnote{Varying the
{\it normalization} of the power-spectrum changes our inferred bias and the
theoretical bias of the dark halos by almost the same factor, and
therefore has little effect on the consistency of our
observations with the dark halo model.  This means that
our conclusions will not be very sensitive to the assumed
cosmological model, as Figure~4 shows.}
but if we restrict ourselves to spectral shapes that are
not grossly inconsistent with local constraints, we find what
is summarized in Figure~4.

Figure~4 shows the linear bias (equation 2) as a function of
spectral shape $\Gamma$ 
for dark halos with abundance equal to the observed abundance
of Lyman-break galaxies.\footnote{The
bias of a population of halos depends upon its
mass distribution, since more massive halos are more strongly clustered
than less massive halos.  We do not know this distribution for
Lyman-break galaxies.
If we assumed the Lyman-break technique detected one galaxy in each halo
more massive than some limit $M_0$, and no galaxies in
halos less massive, we could determine $M_0$ from the abundance
of the galaxies by integrating the
Press-Schechter formula (equation 3).  In fact the technique is likely
to find galaxies in halos less massive than
$M_0$ as well as in halos more massive, and so $M_0$ defined this
way is perhaps close to the typical halo mass.  The bias shown
in Figure~4 is for halos of mass $M_0$, and as such is only an approximation
to the bias of the observed population.  A more sophisticated treatment
will be presented elsewhere.}
Decreasing the amount of small scale power (i.e. decreasing $\Gamma$)
increases the predicted bias of these halos;
the inferred bias of Lyman-break galaxies begins to match the
predicted bias of dark halos
at $\Gamma\sim 0.2$, the locally favored spectral shape, as
would be expected if these galaxies formed within the most massive 
dark matter halos at $z\sim 3$.  At $\Gamma=0.2$ the number density
of Lyman-break galaxies implies typical masses of $6\times 10^{10}$,
$1.4\times 10^{12}$, $8\times 10^{11} h_{100}^{-1} M_{\sun}$ for $\Omega_M=1$,
$\Omega_M=0.2$~open, and $\Omega_M=0.3$~flat.   Though (in the dark-halo model)
measuring the number density and bias of a population of objects
reveals little about cosmology other than the shape of
the power-spectrum, measuring in addition the {\it masses} of the objects
pins down the spectral normalization and provides a sensitive
cosmological probe.
Limited near-infrared spectroscopy on Lyman-break galaxies (Pettini \et 1998)
has so far placed only weak
constraints on their masses; we look forward to the availability
of near-IR spectrographs on 8m-class telescopes.

If it is in fact true that Lyman-break galaxies form within
dark halos, then other conclusions follow from the data.
For example, we have assumed so far that Lyman-break galaxies---the
brightest $z\sim 3$ galaxies in the rest UV---reside
only within the most massive dark halos, but this
need not be true; it is easy to imagine that 
the galaxies brightest in the UV are those with the least
dust, or with the most recent burst of star formation, and that
halo mass is only a secondary consideration.  In
this case there could be a large spread in the UV luminosities
of galaxies within halos of a given mass.  Because
low-mass halos are so much more numerous than high-mass halos,
if the spread were large enough our observed sample would be dominated
by low-mass halos which happened to be UV bright.  The strong
clustering we observe shows that there cannot be
a large population of low-mass (and thus weakly clustered)
interlopers in our sample, and this limits the allowed spread
in UV luminosities for halos of a given mass.  The dotted lines
in Figure~4, showing the bias of halos ten times more abundant
than Lyman-break galaxies, illustrate the point.  These halos
have masses only 4, 8, 5 times lower than halos as abundant
as Lyman-break galaxies for $\Omega_M=1$, $0.2$~open, $0.3$~flat, but if
even 10\% of them contained galaxies bright enough to be included in our
sample the clustering strength would be diluted to well below
what we observe.  The implication is that lower mass halos
are fainter in the UV not just on average, but (nearly) on
a halo-by-halo basis.  
If Lyman break galaxies really were sub-galactic fragments,
rapidly fading after bursts of star formation triggered
by chance interactions with other fragments (e.g. Lowenthal \et 1997),
one might not expect so tight a correlation between
UV luminosity and dark halo mass.
Similar arguments can be used to undermine the claim that
the Lyman-break technique misses a large fraction
of the galaxies in massive halos at $z\sim 3$.\footnote{The opposite
possibility---that there is more than one Lyman-break galaxy per massive
halo---could in principle help reconcile our observations with
standard CDM, but is inconsistent with the small number of close
galaxy pairs in our sample (Giavalisco \et 1998a).}
The uncertainty in the bias is still large, and our
analytic approximations rather crude, so it would be premature
to make too much of arguments such as these; but they show the
kind of conclusions that can be drawn from our sample
in the context of the dark halo model.
These ideas will be developed further elsewhere (Adelberger \et 1998).

\section{SUMMARY}

We have estimated the variance of Lyman-break galaxy counts in cubes 
of side length 7.7, 11.9, 11.4 $h_{100}^{-1}$ Mpc
as $\sigma_{\rm gal}^2\sim 1.3\pm0.4$
for $\Omega_M=1$, $0.2$~open, $0.3$~flat.
This variance implies that Lyman-break galaxies
have a bias $b\equiv\sigma_{\rm gal}/\sigma_{\rm mass}$ of
$b=6.0\pm 1.1$, $1.9\pm 0.4$, $4.0\pm 0.7$ for the same cosmologies.
The bias is in good agreement with a simple model, first
proposed by White \& Rees (1978), in which galaxies form
within virialized halos of dark matter.  The agreement of our data
with this model depends on cosmology primarily through the shape
of the power-spectrum, rather than through the growth rate of
matter perturbations as might have been expected.  Given the abundance
of Lyman-break galaxies and the locally determined power-spectrum shape,
one could have predicted a priori from this model
the clustering strength we have observed.
The agreement is surprisingly good, for it assumes not
only that galaxies form within dark halos---which is plausible
enough---but that galaxies UV-bright enough for us to detect
reside almost exclusively within the most massive halos. 
UV luminosity depends so strongly on the age of a starburst
and on the importance of dust extinction that one might have expected
halo mass to play a comparatively minor role in Lyman-break galaxies'
UV luminosities; but this appears not to be the case.
The observed abundance and clustering properties of Lyman-break
galaxies suggest instead an almost one-to-one
correspondence of massive halos to {\it observable} galaxies, and this
implies, for example, that the most massive halos essentially
always exhibit star formation at detectable levels
(i.e., that the duration of star formation is close to the time interval
over which the galaxies in the sample are observed), and
that halos only slightly less massive rarely do.  The simple analytic
approach adopted in \S 4 cannot justify more precise
statements here; these will be presented elsewhere (Adelberger \et 1998).

While we have argued that our data can be understood through an appealingly
simple model for galaxy formation in which galaxies form within
dark-matter halos, the UV luminosity of young galaxies is tightly correlated
with their mass, and the power-spectrum of mass fluctuations at $z\sim 3$
has a shape similar to that determined locally, this does not of course
rule out other models.  We look forward to learning how well our data
agree with competing scenarios for galaxy formation.  In the meantime,
one prediction of the scenario we favor is that fainter samples of
Lyman-break galaxies in the same redshift range should exhibit
weaker clustering; existing data will allow us to test this
observationally (Giavalisco \et 1998b).

\bigskip
\bigskip

It is a pleasure to acknowledge several conversations with J. Peacock
at the beginning of this project. 
We are grateful to the many people responsible for
building the W.~M.~Keck telescopes and the Low Resolution
Imaging Spectrograph.
Software by J. Cohen, A. Phillips, and P. Shopbell helped
in slit-mask design and alignment.
CCS acknowledges support from the U.S. National Science Foundation through
grant AST 94-57446, and from the Alfred P. Sloan Foundation. 
MG has been supported through grant HF-01071.01-94A from the Space Telescope
Science Institute, which is operated by the Association of Universities for
Research in Astronomy, Inc. under NASA contract NAS 5-26555.

\bigskip
\newpage
\begin{figure}
\figurenum{1}
\plotone{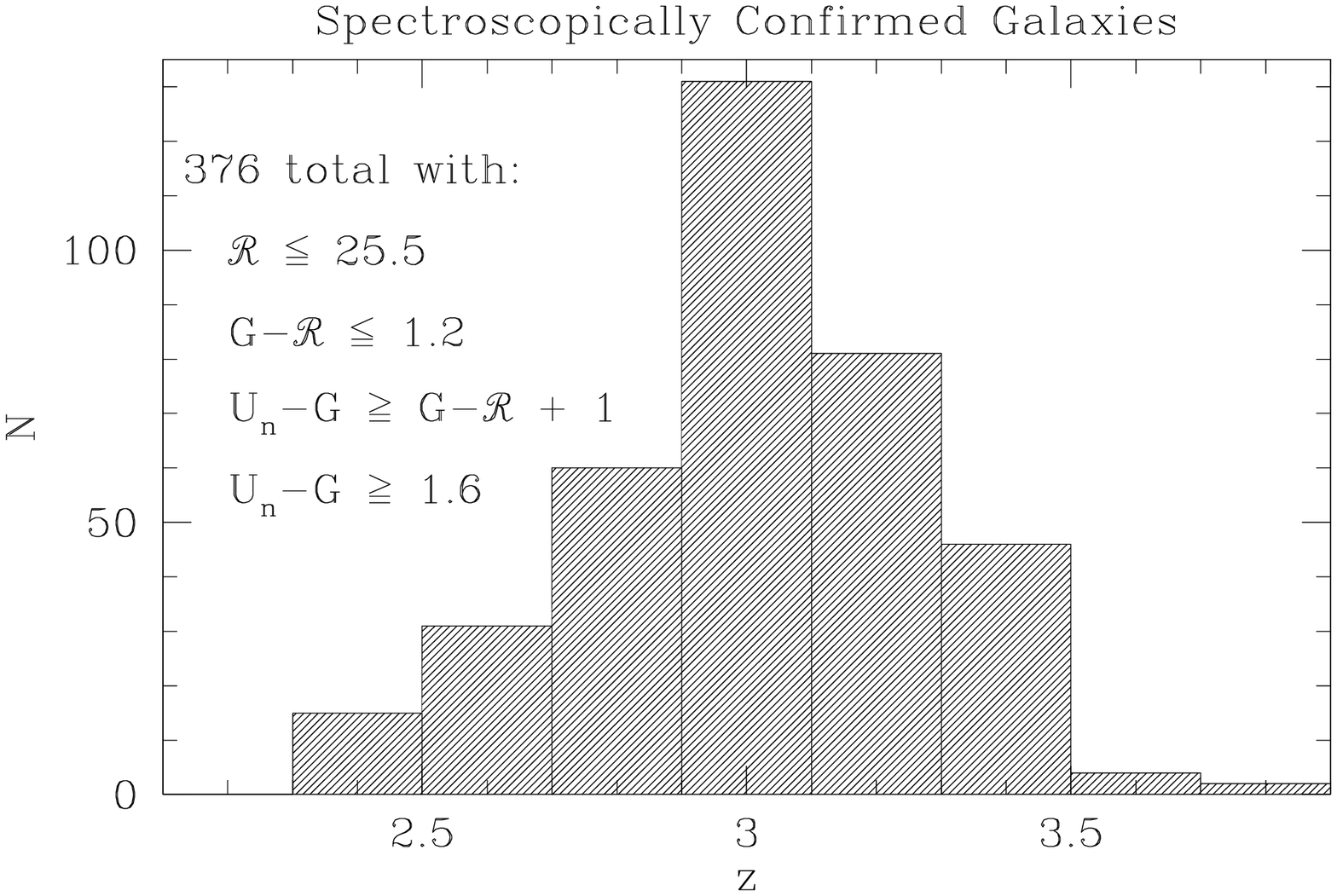}
\figcaption{The redshift distribution of objects satisfying the Lyman-break
criteria adopted in this letter.
Only the 70\% of candidates which
have been confirmed to be galaxies are shown; 25\% of our spectroscopic
sample has not been identified due to low S/N, and 5\% is stars.}
\end{figure}
\newpage
\begin{figure}
\figurenum{2}
\plotone{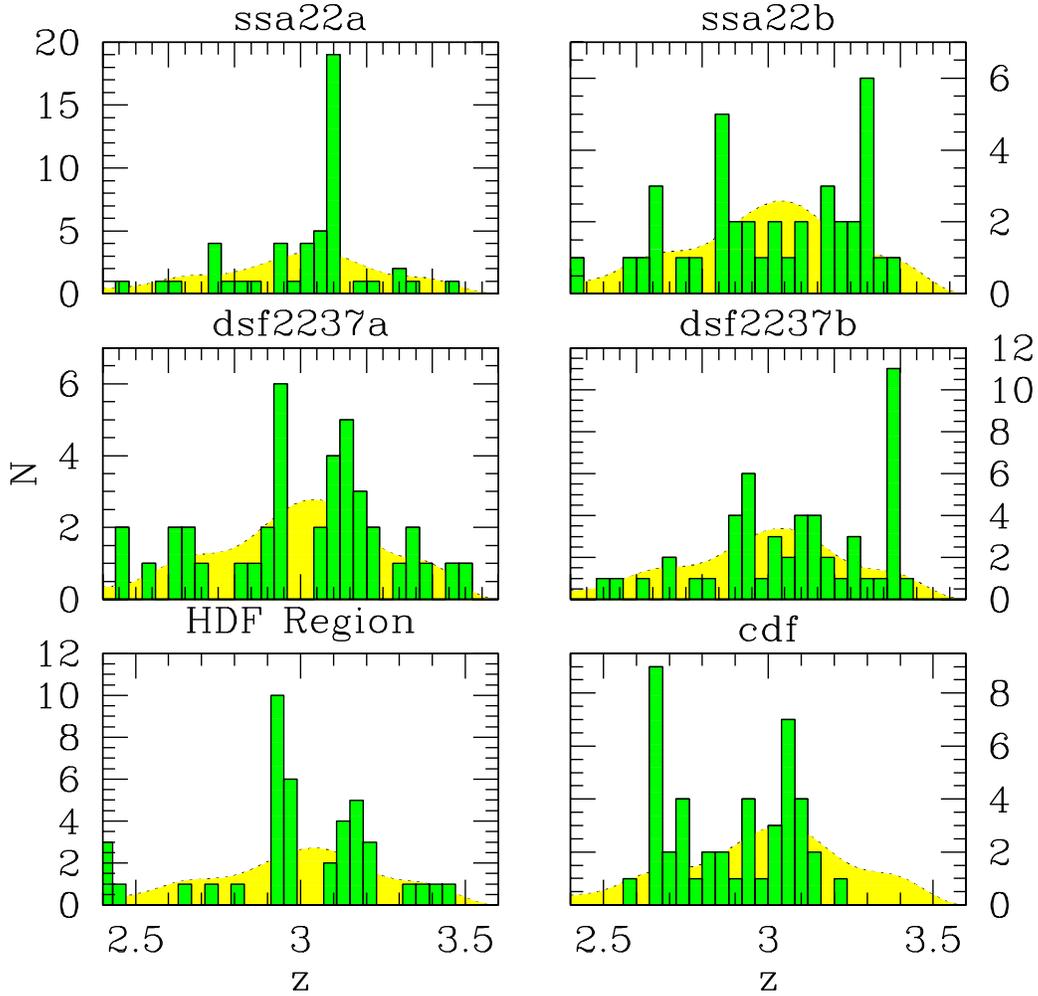}
\figcaption{Redshift distributions in the six fields.  The smooth
curve is our estimated selection function, produced by fitting
a spline to the coarsely binned redshifts of all
candidates matching our selection criteria; the superimposed histograms
are the measured redshifts in each field.  The actual binning used
in our analysis is somewhat different from the binning presented here.
The average number of redshifts per field is about 44, with field to
field variations due mainly to different levels of spectroscopic
completeness.}
\end{figure}
\newpage
\begin{figure}
\figurenum{3}
\plotone{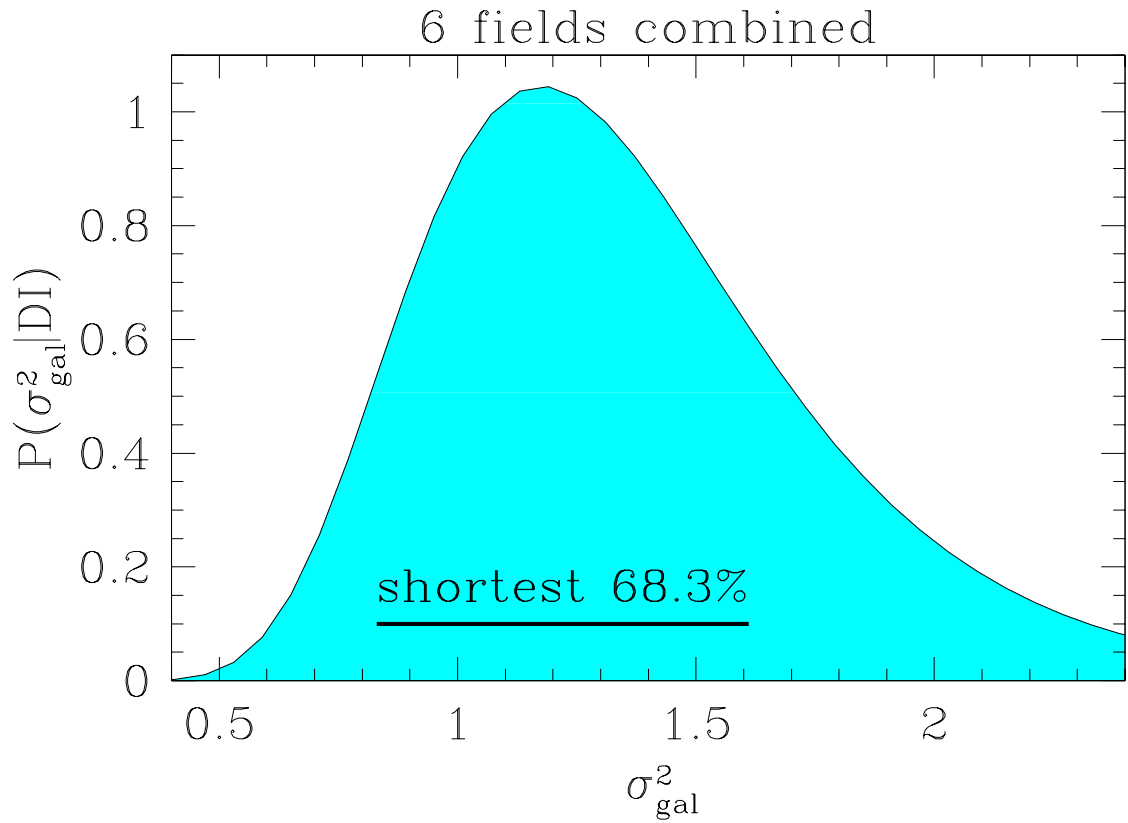}
\figcaption{The posterior probability of our data
as a function of clustering strength $\sigma_{\rm gal}^2$.
A uniform prior in $\sigma_{\rm gal}^2$ is assumed, and the
shortest 68\% credible interval is shown.}
\end{figure}
\newpage
\begin{figure}
\figurenum{4}
\plotone{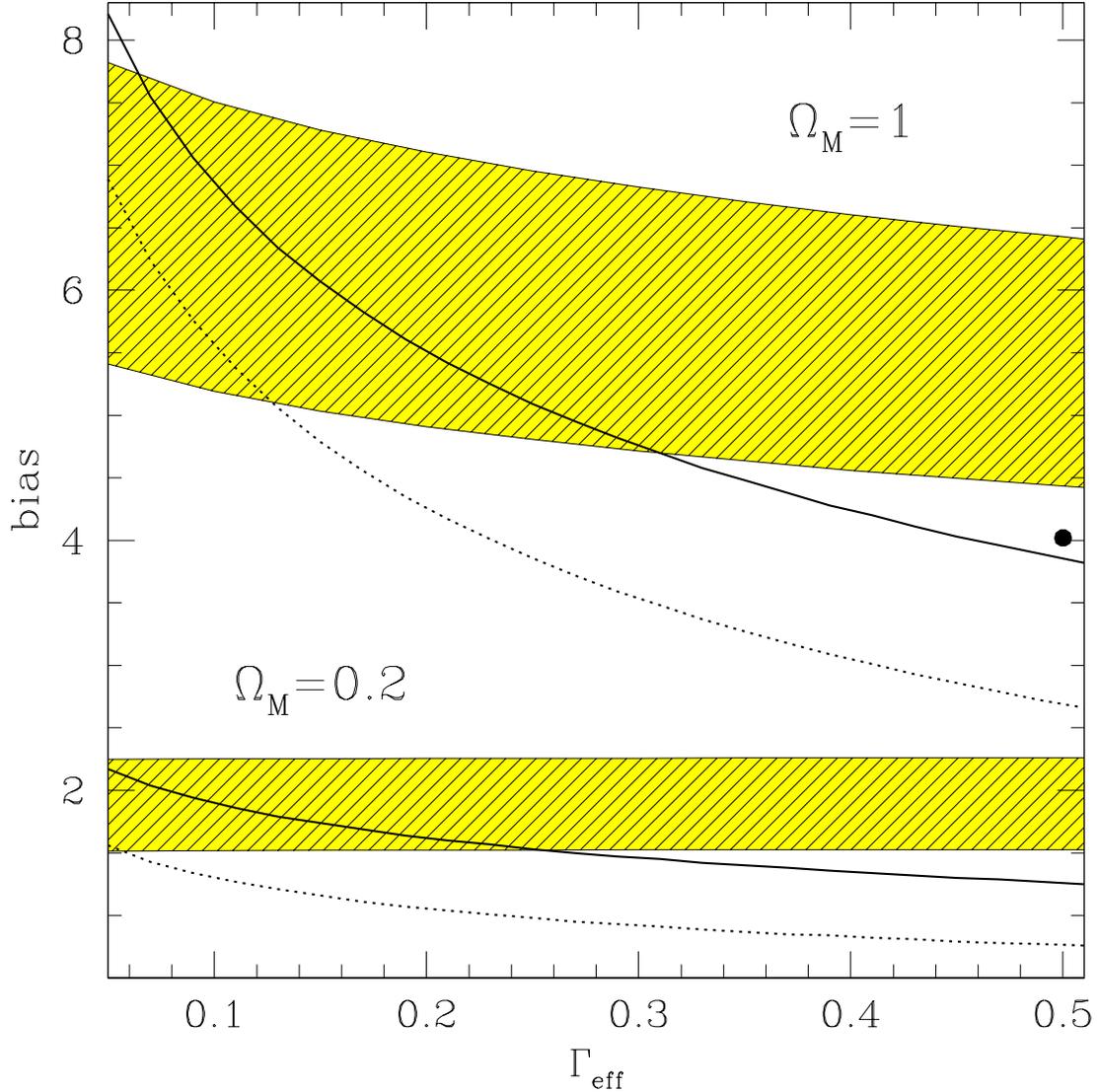}
\figcaption{The spectrum-dependence of bias for objects of fixed abundance.
Rarer (high-$\nu$) halos are more strongly clustered, but the details
of the bias--abundance relationship depend upon the fluctuation power-spectrum.
Spectra with more small scale power (higher $\Gamma$) have a lower
bias for objects of given abundance.  The solid curves show the
predicted bias of dark halos as abundant as Lyman-break galaxies,
and the dotted lines show the bias for halos ten times more abundant.
Our 68\% credible intervals on the bias are shaded.
The results for $\Omega=0.3$~flat, discussed in the text, have been
suppressed for clarity.  The observations
are consistent with all cosmologies considered ($\Omega_M=1$, $0.2$~open,
$0.3$~flat) if the spectral shape $\Gamma$
is treated as a free parameter, though the preferred values $\Gamma\simlt 0.3$
arise more naturally if $\Omega_M<1$.  Standard CDM, with $\Omega_M=1$
and $\Gamma=0.5$, seems to disagree with our data at about the
$2\sigma$ level---but this is hardly the worst of its problems.
The analytic approximations used are rather crude, and
the point to the right of the plot, drawn from the
$\Gamma=0.5, \Omega_M=1$ N-body simulation of Jing \& Suto (1998),
gives some idea of their reliability.  The N-body estimate
of the bias has been scaled to the values of $\sigma_8$ we adopt 
($\sigma_8=0.5$ for $\Omega_M=1$, $\sigma_8=1$ for $\Omega_M=0.2$).}
\end{figure}
\end{document}